\def\beq{\begin{equation}}
\def\enq{\end{equation}}

\documentclass{PoS}

\usepackage{aas_macros}

\usepackage{amsmath}
\usepackage{amssymb}

\title{Effective action for the Abelian-Higgs model for a gauge-invariant implementation on optical lattices}

\ShortTitle{Effective action for the Abelian-Higgs model on optical lattices}

\author{\speaker{Alexei Bazavov}$^{a,b}$, Yannick Meurice$^b$, Shan-Wen Tsai$^a$,
Judah Unmuth-Yockey$^b$, Jin Zhang$^a$\\
\llap{$^a$}Department of Physics and Astronomy, University of California, Riverside, CA 92521, USA\\
\llap{$^b$}Department of Physics and Astronomy, University of Iowa, Iowa City, IA 52242, USA\\
        E-mail: \email{obazavov@quark.phy.bnl.gov}
}

\abstract{
We present a gauge-invariant effective action for the Abelian-Higgs model in 1+1 dimensions. It is constructed by integrating out the gauge field and then using the hopping parameter expansion. The latter is tested with Monte Carlo simulations for small values of the scalar self-coupling. In the opposite limit, at infinitely large self-coupling, the Higgs mode is frozen and the partition function can be written in terms of local tensors and the tensor renormalization group blocking can be applied. The numerical implementation requires truncations and the time continuum limit of the blocked transfer matrix can be obtained numerically. At zero gauge coupling and with a spin-1 truncation, the small volume energy spectrum is identical to the low energy spectrum of a two-species Bose-Hubbard model in the limit of large onsite repulsion. The procedure is extended to finite gauge coupling and we derive a spin-1 approximation of the Hamiltonian which involves terms corresponding to transitions among the two species in the Bose-Hubbard model. An optical lattice implementation involving a ladder structure is proposed.
}

\FullConference{The 33rd International Symposium on Lattice Field Theory\\
		 14 -18 July  2015\\
		 Kobe International Conference Center, Kobe, Japan}

\begin{document}

\section{Introduction}
\label{sec:intro}

There has been a lot of interest recently in the possibility
of building quantum simulators for lattice gauge theory
using optical lattices~\cite{Wiese:2013uua,Zohar:2015hwa}.
When gauge fields are mapped onto the physical degrees of freedom,
such as individual atoms or condensates trapped in an optical lattice,
one needs to ensure that the physical observables are gauge invariant.
This can be achieved by imposing Gauss's law (which has often
to be done approximately)
or by constructing an explicitly gauge invariant formulation of the
original model. We report on our recent progress in following
the latter approach~\cite{PhysRevD.92.076003}.

We establish an approximate {\it quantitative} correspondence between the
Abelian-Higgs model (scalar electrodynamics)
on a 1+1 dimensional  lattice and a two-species Bose-Hubbard model with
specifically chosen interactions that can in principle be realized
experimentally on optical lattices.
We derive the gauge-invariant effective action
by integrating over the gauge fields.
In this case the constraint enforcing gauge invariance
is satisfied automatically.

\section{The Abelian-Higgs model and the effective action}
\label{sec:abh}

We consider the Abelian-Higgs model
on a 1+1 space-time lattice of size $N_s\times N_\tau$.
The scalar field $\phi_x=|\phi_x|\exp(i\theta_x)$ is attached to the sites
and the gauge fields 
$U_{x,\hat{\nu}}=\exp(i A_{x,\hat\nu})$ to the links of the lattice.
The product of $U$'s around a plaquette is denoted $U_{pl,x}$ where $x$ is the lower left corner of the plaquette in space-time coordinates. $\beta_{pl}=1/g^2$ is the inverse gauge coupling and $\kappa_s$ ($\kappa_\tau$) is the hopping coefficient in the space (time) direction. 
The partition function and the action are:
\begin{eqnarray}\label{S_u1h}
Z &=& \int D\phi^\dagger D\phi DU e^{-S},\,\,\,\,\,\,\,\,\,
S = -\beta_{pl}\sum_x{\rm Re}\left[U_{pl,x}\right]
+\sum_x\phi_x^\dagger\phi_x
+\lambda\sum_x\left(\phi_x^\dagger\phi_x-1\right)^2\nonumber\\
&-& {\kappa_\tau}\sum_x
\left[{\rm e}^{\mu}\phi_x^\dagger U_{x,\hat\tau}\phi_{x+\hat\tau}+{\rm e}^{-\mu}
\phi_{x+\hat\tau}^\dagger U^\dagger_{x,\hat\tau}\phi_x
\right] -{\kappa_s}\sum_x
\left[\phi_x^\dagger U_{x,\hat{s}}\phi_{x+\hat{s}}+
\phi_{x+\hat{s}}^\dagger U^\dagger_{x,\hat{s}}\phi_x
\right].
\end{eqnarray}
The Nambu-Goldstone fields $\theta_x$ can be eliminated by a gauge transformation 
\begin{equation}
A_{x,\hat\nu}\rightarrow A_{x,\hat\nu}-\theta_{x+\hat\nu} + \theta_{x},
\end{equation}
which leaves the plaquette terms unchanged.

Unlike other approaches,
\textit{e.g.}~\cite{Wiese:2013uua,Zohar:2015hwa,Tagliacozzo:2012vg,Zohar:2011cw,Zohar:2012ay,Kasamatsu:2012im,Kuno:2014npa},
we will not try to implement the gauge field on the optical lattice, but rather try to implement a gauge-invariant effective action obtained by integrating over the gauge fields.
A similar approach is being pursued for theories with fermions~\cite{Vairinhos:2014uxa}.
For this purpose, we use the Fourier expansion of the Boltzmann weights in terms of the modified Bessel functions $I_n$, for instance, 
\begin{eqnarray}
&\ &\exp[2\kappa_\tau |\phi_x||\phi_{x+\hat\tau}|  \cos(\theta_{x+\hat\tau} - \theta_{x}+A_{x,\hat\tau}-i\mu)]\label{eq_dual}\\
&=&\sum_{n=-\infty} ^{\infty}I_n(2\kappa_\tau |\phi_x||\phi_{x+\hat\tau}|)\exp[in(\theta_{x+\hat\tau} - \theta_{x}+A_{x,\hat\tau}-i\mu)], \nonumber \end{eqnarray}
and similar expressions for the space hopping and the plaquette interactions. We can then collect 
all the exponentials involving a given $A_{x,\hat\nu}$ and perform the integration over $A_{x,\hat\nu}$. This results in Kronecker deltas relating the various Fourier modes. The final result is 
\begin{equation}
e^{-S_{eff}}=\sum_{ \{ m_\Box \} }
\left[
\prod_{\Box} I_{m_{\Box}}(\beta_{pl})
\prod_{x}
\bigg(
I_{n_{x,\hat{s}}}(2\kappa_s |\phi_x||\phi_{x+\hat{s}}| )
I_{n_{x,\hat{\tau}}}(2\kappa_\tau |\phi_x||\phi_{x+\hat{\tau}}| )
\exp(\mu n_{x,\hat{\tau}})\bigg)
\right],
\label{eq_Seff}
\end{equation}
where the link indices $n_{x,\hat\nu}$ are related to
the plaquette indices $m_\Box$ by the rules
\begin{equation}
n_{x,\hat{s}} = m_{below}-m_{above},\,\,\,\,\,\,\,\,
n_{x,\hat\tau} = m_{right}-m_{left},
\label{eq:nandm2}
\end{equation}
where the subscripts of $m$s refer to the plaquette location with respect to the link.
Eqs. (\ref{eq:nandm2}) guarantee that the link indices automatically  satisfy the current
conservation imposed by the integration of the $\theta_x$ variables, so
the $m_\Box$ act as the dual variables.

Equations (\ref{eq:nandm2}) have simple electromagnetic analogs.
First, $n_{x,\hat\tau}$ can be interpreted as a charge and $m_x$
as an electric field in the spatial direction.
With this Minkowskian interpretation, Eq.~(\ref{eq:nandm2}) enforces Gauss's law. Second, $n_{x,\hat\nu}$ can be interpreted as a two-dimensional current and $m_x$ as a magnetic field normal 
to the two-dimensional plane. In this Euclidean interpretation, Eqs. (\ref{eq:nandm2}) express the current as the curl of the magnetic field.

At the lowest order of the strong-coupling expansion we have $\beta_{pl}=0$ and from $I_n(0)=0$ for $n\neq 0$, we see that all the indices must be zeros. 
The effect of the plaquette can be restored perturbatively. This can be organized in an expansion in the 
hopping parameter. In the isotropic case $\kappa_\tau=\kappa_s=\kappa$ we obtain: 
\begin{equation}
S_{eff}=\sum_{\langle xy\rangle}\left(-\kappa^2 M_xM_y+\frac{1}{4}\kappa^4(M_xM_y)^2\right)
-2\kappa^4\frac{I_1(\beta_{pl})}{I_0(\beta_{pl})}\sum_{\Box (xyzw) }M_xM_yM_zM_w +O(\kappa^6),
\end{equation}
where $M_x=\phi^\dagger_x\phi_x$ is a composite gauge-invariant field.

\section{Monte Carlo calculations}
\label{sec:MC}

We consider the action (\ref{S_u1h}) for the isotropic case $\kappa_\tau=\kappa_s=\kappa$.
In the $\beta_{pl}\to\infty$ limit when $U_{x,\hat{\nu}}=1$, 
the expectation value of the hopping term
$L_\phi = \langle{\rm Re}\{\phi_x^\dagger U_{x,\hat{\nu}}\phi_{x+\hat{\nu}}\}\rangle$
can be calculated with the hopping parameter expansion, for small $\kappa$.
It has been derived up to $O(\kappa^5)$ in Ref.~\cite{heitgerphd}. This result
can be generalized to $\beta_{pl}<\infty$ by including the appropriate factors
of $I_1(\beta_{pl})/I_0(\beta_{pl})$ for the diagrams that involve plaquettes.

\begin{figure}
\begin{center}
\includegraphics[width=0.495\textwidth]{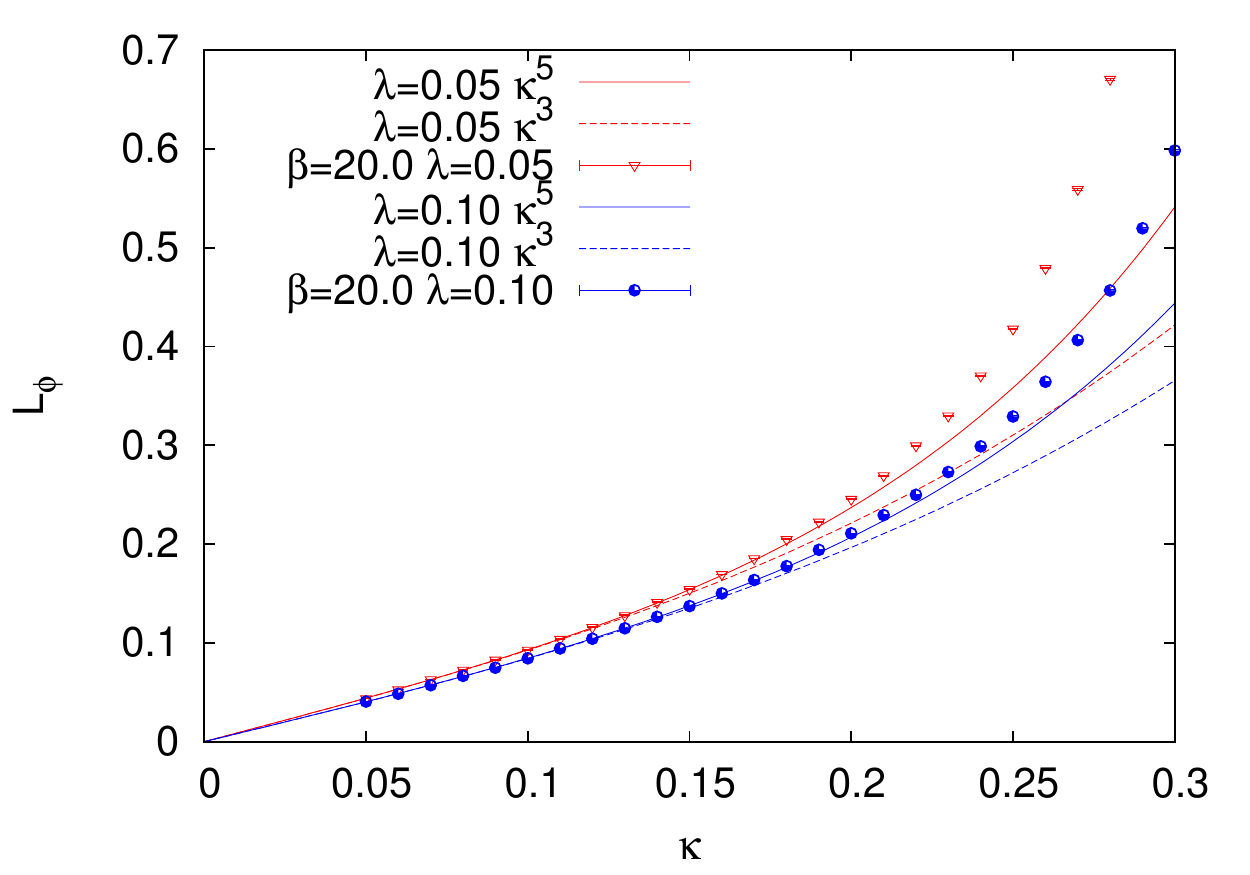}\hfill
\includegraphics[width=0.495\textwidth]{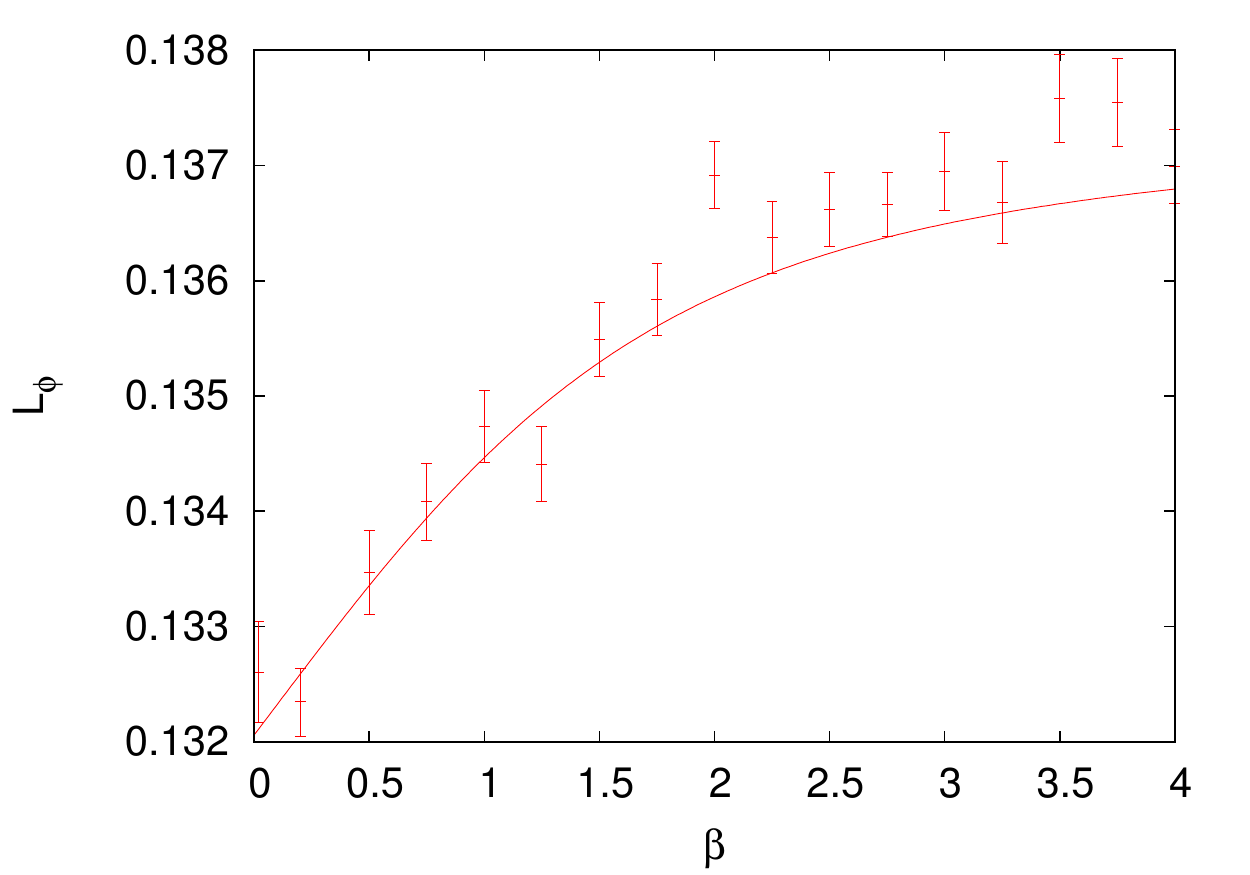}
\parbox[b]{0.48\textwidth}{
\caption{
$L_\phi$ at $\beta_{pl}=20$ for $\lambda=0.05$ and $\lambda=0.1$ as function
of $\kappa$ compared with the
hopping expansion at $\beta_{pl}=\infty$ at $O(\kappa^3)$ and $O(\kappa^5)$.
}\label{fig_Lf_inf}
}
\hfill
\parbox[b]{0.48\textwidth}{
\caption{
$L_\phi$ at fixed $\kappa=0.15$ as function of $\beta_{pl}$ for $\lambda=0.1$ compared with the
hopping expansion with included dependence on $\beta_{pl}$ up to $O(\kappa^5)$.
}
\label{fig_Lf_kappa}
}
\end{center}
\end{figure}

To check the range of validity of the expansion we perform Monte Carlo
simulations at several values of $\beta_{pl}$, $\kappa$ and $\lambda$ on
a $16^2$ lattice. To test the $\beta_{pl}\to\infty$ limit we set $\beta_{pl}=20$ and
for $\lambda=0.05$ and $0.1$ scan the range of $\kappa\in[0.05,0.30]$.
The results for $L_\phi$ are shown in Fig.~\ref{fig_Lf_inf}. The lines represent
the expansion at two orders. The expansion starts to break down around $\kappa=0.15$
at $O(\kappa^3)$ and $\kappa=0.2$ at $O(\kappa^5)$.

To understand the dependence on $\beta_{pl}$ better we also calculate $L_\phi$ for several
values of $\beta_{pl}$ at fixed $\kappa=0.15$. The results together with the
hopping expansion are shown in Fig.~\ref{fig_Lf_kappa}.
There is good agreement, and the dependence on $\beta$ is weak.

\section{The large $\lambda$ limit}
\label{sec:lambda}
 When $\lambda$ becomes arbitrarily large, $M_x$ is frozen to 1, or in other words, the 
 Brout-Englert-Higgs mode becomes infinitely massive. We are then left with compact variables of integration in the original formulation ($\theta_x$ and $A_{x,\hat\nu}$) and the Fourier expansions described before lead to expressions of the partition function in terms of discrete sums. 
 
 As explained in Ref. \cite{PhysRevD.88.056005}, these sums can be formulated in a compact way using tensorial notations.  The partition function can be written as
\begin{equation}
Z= ( I_{0}(\beta_{pl})I_0(2\kappa_s)I_0(2\kappa_\tau))^V
{\rm Tr}
\left[
\prod_{s,\tau,\Box}A^{(s)} _{m_{above}m_{below}}
A^{(\tau)} _{m_{left}m_{right}}
B^{(\Box)}_{m_1m_2m_3m_4}
\right],
\end{equation}
where the $A$ tensors are associated with the spatial and temporal links and $B$ tensors
with the plaquettes. $A$ and $B$ tensors are defined
in terms of the ratios of Bessel functions $t_{n}(z) = I_{n}(z)/I_{0}(z)$,
see Ref.~\cite{PhysRevD.92.076003} for details.

The traces can also be expressed in terms of a transfer matrix $\mathbb{T}$ which can be constructed
by defining the following tensors:
 \begin{eqnarray}
\mathbb{B}_{(m_1,m_2,\dots m_{N_s})(m_1',m_2'\dotsm_{N_s}')} &=&
t_{m_1}(2\kappa_\tau)\delta_{m_1,m_1'}t_{m_1}(\beta_{pl})
t_{|m_1-m_2|}(2\kappa_\tau)\delta_{m_2,m_2'}t_{m_2}(\beta_{pl})\nonumber\\
&\times&t_{|m_2-m_3|}(2\kappa_\tau)\dots
t_{m_{N_s}}(\beta_{pl})t_{m_{N_s}}(2\kappa_\tau).\\
\mathbb{A}_{(m_1,m_2,\dots m_{N_s})(m_1',m_2'\dots m_{N_s}')} &=&
t_{|m_1-m_1'|}(2\kappa_s)t_{|m_2-m_2'|}(2\kappa_s)\dots t_{|m_{N_s}-m_{N_s}'|}(2\kappa_s)
\end{eqnarray}
(this definition implies open boundary conditions, $m=0$ at both ends).
In this case, the chemical potential has completely disappeared. If we had chosen different $m$'s at the end 
allowing a total charge $Q$ inside the interval, we would have an additional factor $\exp (\mu Q)$.

Since $\mathbb{B}$ is diagonal, real and positive,
the symmetric transfer matrix can be defined as
\beq
\label{eq:tm}\mathbb{T}=\sqrt{\mathbb{B}}\mathbb{A}\sqrt{\mathbb{B}}\,\,\,\,\,\,\,\,
\mbox{and}\,\,\,\,\,\,\,\,
Z= ( I_{0}(\beta_{pl})I_0(2\kappa_s)I_0(2\kappa_\tau))^V {\rm Tr}\left[\mathbb{T}^{N_\tau}\right].
\enq
The $\mathbb{A}$ and $\mathbb{B}$ matrices can be constructed by a recursive blocking method similar 
to those discussed in Ref. \cite{PhysRevD.88.056005}.

In the case where both $\lambda$ and $\beta_{pl}$ are infinite this
model corresponds to the two-dimensional classical $O(2)$ model.
A Hamiltonian formulation of the model suitable for quantum simulation
on optical lattice has been proposed by some of the present authors
in Ref.~\cite{PhysRevA.90.063603}.

\section{The time continuum limit, energy spectrum
and Bose-Hubbard model implementation}
\label{sec:time}

We construct the time continuum limit of the transfer matrix $\mathbb{T}$ by
taking the limit $1<<\beta_{pl}<<\kappa_\tau$ and
keeping the open boundary conditions ($m=0$ at both ends).
At leading order in the inverse of these large parameters, the eigenvalues of $\mathbb{T}$ are
\begin{eqnarray}
\label{eq:eig}
\lambda_{(m_1,m_2,\dots m_{N_s})} &=&
1-\frac{1}{2}\left[\frac{1}{\beta_{pl}} (m_1^2+m_2^2+\ldots+m_{N_s}^2)\right.
\nonumber\\
&+&\left.\frac{1}{2\kappa_\tau}(m_1^2+(m_2-m_1)^2+\ldots
 +(m_{N_s}-m_{N_s-1})^2+m_{N_s}^2)\right]
\end{eqnarray}

We set the scale with the gap energy $\tilde{U}_P$ and relate it to the other
energy scales as:
\beq
\tilde{U}_P\equiv \frac{1}{a\beta_{pl}},\,\,\,\,\,
\tilde{Y}\equiv \frac{1}{2\kappa_\tau a} =\frac{\beta_{pl}}{2\kappa_\tau} \tilde{U}_P,\,\,\,\,\,
\tilde{X}\equiv\sqrt{2}\beta_{pl}\kappa_s\tilde{U}_P.
\enq

We can now derive an expression for the Hamiltonian in the spin-1 approximation
where the plaquette quantum number $m$ takes values $\pm1$ and 0 only.
The effect of $\kappa_s$ can be included by linearization.
The final form of the Hamiltonian $\bar{H}$ for $1<<\beta_{pl}<<\kappa_\tau$ is 
\begin{equation}
\label{eq:ham}
\bar{H}=\frac{\tilde{U}_P}{2}\sum_i \left(\bar{L}^z_{(i)}\right)^2+
\frac{\tilde{Y}}{2} {\sum_i}  ' (\bar{L}^z_{(i)}-\bar{L}^z_{(i+1)})^2-
\tilde{X}\sum_{i}
\bar L^x_{(i)}\ ,
\end{equation}
where $\sum_i '$ is a short notation to include the single terms at the two ends as in Eq. (\ref{eq:eig}), \textit{i.e.} besides $(\bar{L}^z_{(1)}-\bar{L}^z_{(2)})^2$, $(\bar{L}^z_{(2)}-\bar{L}^z_{(3)})^2$, ... , $(\bar{L}^z_{(N_s-1)}-\bar{L}^z_{(N_s)})^2$ terms this sum contains $(\bar{L}^z_{(1)})^2$ and $(\bar{L}^z_{(N_s)})^2$.

A generic two-species Bose-Hubbard Hamiltonian for a 
linear optical lattice realization of the Abelian-Higgs model
has the form~\cite{PhysRevD.92.076003}:
\begin{eqnarray}
\label{eq_H2}
&\mathcal{H}&=-\sum_{\langle ij\rangle}(t_a a^\dagger_i a_j+t_b b^\dagger_i b_j+h.c.)-\sum_{i, \alpha}(\mu_{a+b}+\Delta_\alpha)n^\alpha_i\nonumber\\
&+&\sum_{i, \alpha}\frac{U_\alpha}{2}n^\alpha_i(n^\alpha_i-1)+W\sum_in^a_in^b_i+\sum_{\langle ij\rangle\alpha}V_\alpha n^\alpha_in^\alpha_j
-\frac{t_{ab}}{2}\sum_i  (a^\dagger_i b_i+  b^\dagger_i a_i)
\end{eqnarray}
with  $\alpha=a,b$ indicating two different species and
with $n^a_i=a^\dagger_ia_i$ and $n^b_i=b^\dagger_ib_i$.
In the limit where $U_a=U_b=U$ and $W$ and $\mu_{a+b}=(3/2) U$ are much larger than any other energy scale, 
we have the condition $n^a_i+n^b_i=2$
for the low energy sector. The three states 
$|2,0\rangle$, $|1,1\rangle$ and $|0,2\rangle$ satisfy this condition
and correspond to the three states of the spin-1 projection considered above.

At the second-order
degenerate perturbation theory this Hamiltonian can be rewritten in terms of the angular
momentum operators:
\begin{eqnarray}
\label{eq:H_eff}
\mathcal{H}_{eff}&=&
\left(\frac{V_a}{2}-\frac{t_a^2}{U}+\frac{V_b}{2}-\frac{t_b^2}{U}\right)\sum_{\langle ij\rangle}L^z_iL^z_j
+\frac{-t_at_b}{U}\sum_{\langle ij\rangle}(L^+_iL^-_j+L^-_iL^+_j)+(U-W)\sum_{i}(L^z_i)^2\nonumber\\
&+&\left[\left(\frac{pn}{2}V_a+\Delta_a-\frac{p(n+1)t_a^2}{U}\right)
-\left(\frac{pn}{2}V_b
+\Delta_b-\frac{p(n+1)t_b^2}{U}\right)\right]\sum_{i}L^z_i
-t_{ab}\sum_{i}L^x_{i},
\end{eqnarray}
where $p$ is the number of neighbors and $n$ is the occupation ($p=2$, $n=2$ in the case under consideration).
$\hat L$ is the angular momentum operator in the representation $n/2$.

\begin{figure}[h]
\centering
 \includegraphics[width=0.40\textwidth]{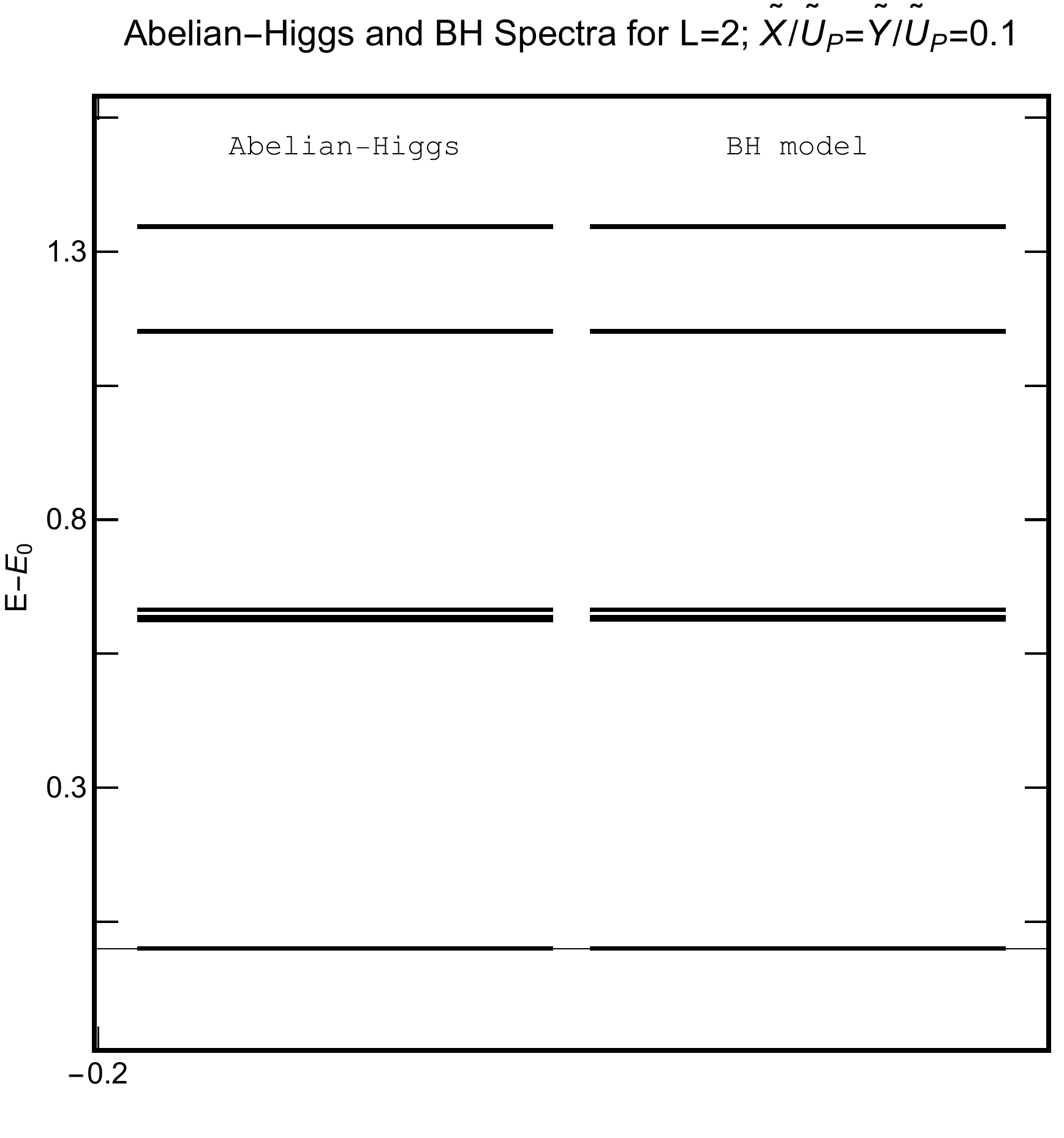}\hspace{1cm}
  \includegraphics[width=0.40\textwidth]{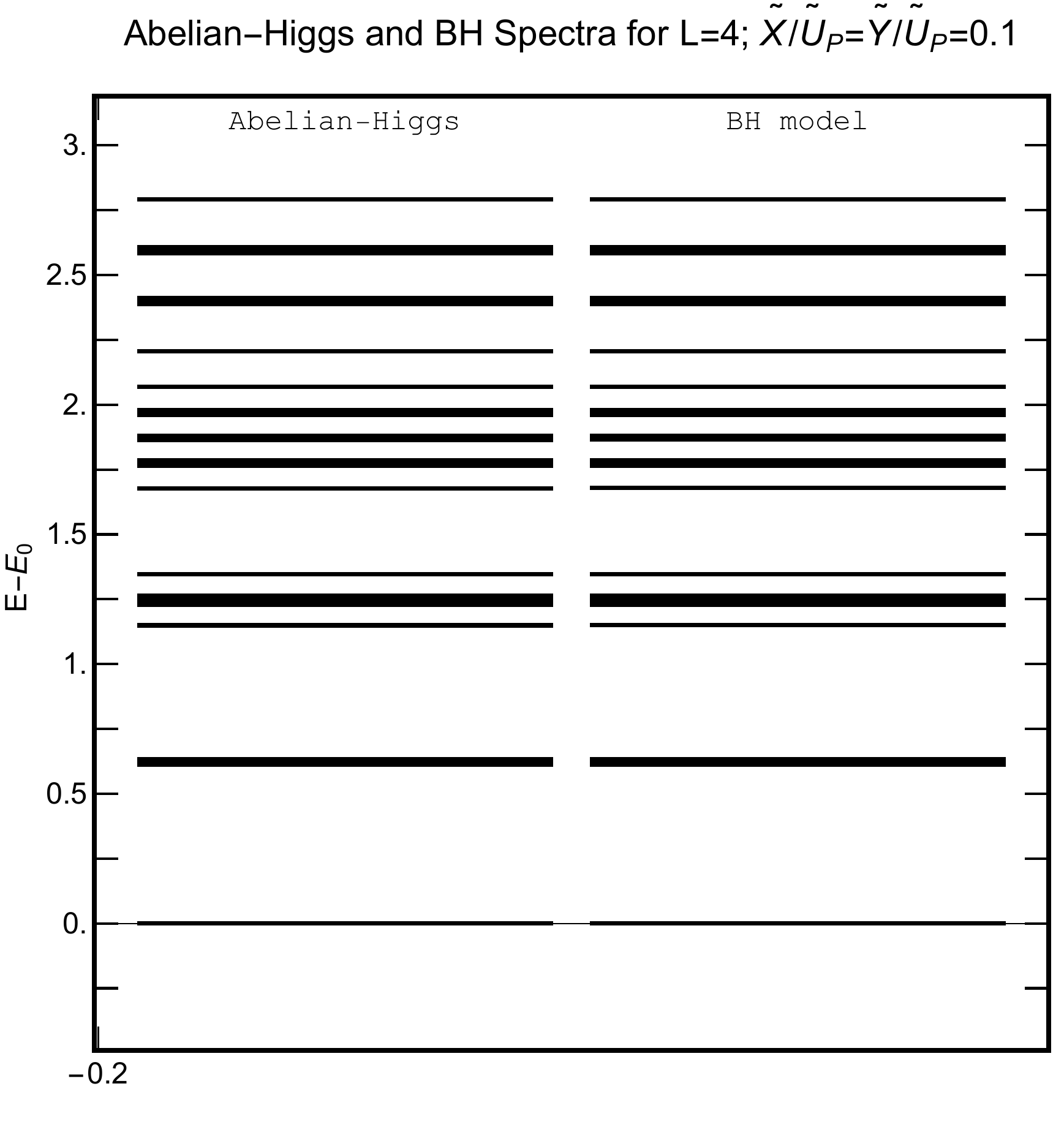}
\caption{\label{fig:hbh4} Abelian-Higgs model with $\tilde{X}/\tilde{U}_P =0.1$, $\tilde{Y}/\tilde{U}_P =0.1$ and the corresponding Bose-Hubbard spectra for $L=2$ (left) and $L=4$ (right).}
\end{figure}

By imposing $t_a=t_b=0$, $V_a=V_b=-\tilde{Y}$/2 and $t_{ab}=\tilde{X}$ we match
the two-species Bose-Hubbard Hamiltonian (\ref{eq:H_eff}) to the one given in Eq.~(\ref{eq:ham}).
In Fig.~\ref{fig:hbh4} we show the spectra of the Abelian-Higgs model obtained with
the tensor renormalization group method
and the two-species Bose-Hubbard model obtained with exact diagonalization
for a system of two (left) and four (right) sites.

The last term in the Hamiltonian, Eq. (\ref{eq_H2}), interchanges
the species index and thus rules out implementing this Hamiltonian
using mixtures of two different types of atoms.
It could be realized with a single atomic species on a ladder structure with $a$ and $b$ corresponding to the two legs of the ladder.
For the experimental implementation, an attractive nearest neighbor interaction ($V_a=V_b=-\tilde{Y}/2$) can be obtained by using cold dipolar atoms or molecules, with dipole moments aligned along the ladder and with inter-rung distance such that the rapidly decaying dipole-dipole interaction between next-nearest-neighbors can be neglected.
Alternatively, the two boson species in the Bose-Hubbard Hamiltonian could correspond to two hyperfine states of the same atom species, provided these two states support Raman transitions between them so that the species conversion term can be created.

\section{Conclusions}

With the remarkable experimental progress in the field of
ultra-cold atoms trapped in optical lattices, and the
unprecedented levels of control and tunability of their
interactions, a wide range of quantum many-body lattice
Hamiltonians can be realized and studied.
We have proposed a two-species Bose-Hubbard model that may be used
as a quantum  simulator for the Abelian-Higgs model in the case of
frozen radial mode in 1+1 dimensions.
The gauge invariance is built-in and thus does not need to be
achieved via fine-tuning, and the correspondence between the proposed
Bose-Hubbard Hamiltonian and the Abelian-Higgs model can be checked quantitatively.
We demonstrate the matching between the two models by calculating
the spectra for small systems.

\acknowledgments
This research was supported in part  by the Department of Energy
under Award Numbers DOE grant DE-FG02-05ER41368, DE-SC0010114 and DE-FG02-91ER40664, the NSF under grants DMR-1411345
and PHY11-25915 and by the Army Research Office of the Department of Defense under Award Number W911NF-13-1-0119.
We also acknowledge helpful discussions with the participants of the
workshops
INT-15-1 ``Frontiers in Quantum Simulation with Cold Atoms" at INT, Seattle, 
``Understanding Strongly Coupled Systems
in High Energy and Condensed Matter Physics'' and
``Ultra-Cold Quantum Matter with Atoms and Molecules''
at the Aspen Center for Physics.

\providecommand{\href}[2]{#2}\begingroup\raggedright\endgroup


\begin{thebibliography}{10}

\bibitem{Wiese:2013uua}
U.-J. Wiese, {\it {Ultracold Quantum Gases and Lattice Systems: Quantum
  Simulation of Lattice Gauge Theories}},  {\em Annalen Phys.} {\bf 525} (2013)
  777--796, [\href{http://arxiv.org/abs/1305.1602}{{\tt arXiv:1305.1602}}].

\bibitem{Zohar:2015hwa}
E.~Zohar, J.~I. Cirac, and B.~Reznik, {\it {Quantum Simulations of Lattice
  Gauge Theories using Ultracold Atoms in Optical Lattices}},
  \href{http://arxiv.org/abs/1503.02312}{{\tt arXiv:1503.02312}}.

\bibitem{PhysRevD.92.076003}
A.~Bazavov, Y.~Meurice, S.-W. Tsai, J.~Unmuth-Yockey, and J.~Zhang, {\it
  Gauge-invariant implementation of the abelian-higgs model on optical
  lattices},  {\em Phys. Rev. D} {\bf 92} (Oct, 2015) 076003.

\bibitem{Tagliacozzo:2012vg}
L.~Tagliacozzo, A.~Celi, A.~Zamora, and M.~Lewenstein, {\it {Optical Abelian
  Lattice Gauge Theories}},  {\em Annals Phys.} {\bf 330} (2013) 160--191,
  [\href{http://arxiv.org/abs/1205.0496}{{\tt arXiv:1205.0496}}].

\bibitem{Zohar:2011cw}
E.~Zohar and B.~Reznik, {\it {Confinement and lattice QED electric flux-tubes
  simulated with ultracold atoms}},  {\em Phys.Rev.Lett.} {\bf 107} (2011)
  275301, [\href{http://arxiv.org/abs/1108.1562}{{\tt arXiv:1108.1562}}].

\bibitem{Zohar:2012ay}
E.~Zohar, J.~I. Cirac, and B.~Reznik, {\it {Simulating Compact Quantum
  Electrodynamics with ultracold atoms: Probing confinement and nonperturbative
  effects}},  {\em Phys.Rev.Lett.} {\bf 109} (2012) 125302,
  [\href{http://arxiv.org/abs/1204.6574}{{\tt arXiv:1204.6574}}].

\bibitem{Kasamatsu:2012im}
K.~Kasamatsu, I.~Ichinose, and T.~Matsui, {\it {Atomic Quantum Simulation of
  the Lattice Gauge-Higgs Model: Higgs Couplings and Emergence of Exact Local
  Gauge Symmetry}},  {\em Phys.Rev.Lett.} {\bf 111} (2013), no.~11 115303,
  [\href{http://arxiv.org/abs/1212.4952}{{\tt arXiv:1212.4952}}].

\bibitem{Kuno:2014npa}
Y.~Kuno, K.~Kasamatsu, Y.~Takahashi, I.~Ichinose, and T.~Matsui, {\it
  {Real-time dynamics and proposal for feasible experiments of lattice
  gauge-Higgs model simulated by cold atoms}},  {\em New J. Phys.} {\bf 17}
  (2015), no.~6 063005, [\href{http://arxiv.org/abs/1412.7605}{{\tt
  arXiv:1412.7605}}].

\bibitem{Vairinhos:2014uxa}
H.~Vairinhos and P.~de~Forcrand, {\it {Lattice gauge theory without link
  variables}},  {\em JHEP} {\bf 12} (2014) 038,
  [\href{http://arxiv.org/abs/1409.8442}{{\tt arXiv:1409.8442}}].

\bibitem{heitgerphd}
J.~Heitger, {\it Numerical simulations of gauge-higgs models on the lattice},
  1997.
\newblock (Ph. D. thesis).

\bibitem{PhysRevD.88.056005}
Y.~Liu, Y.~Meurice, M.~P. Qin, J.~Unmuth-Yockey, T.~Xiang, Z.~Y. Xie, J.~F. Yu,
  and H.~Zou, {\it Exact blocking formulas for spin and gauge models},  {\em
  Phys. Rev. D} {\bf 88} (Sep, 2013) 056005.

\bibitem{PhysRevA.90.063603}
H.~Zou, Y.~Liu, C.-Y. Lai, J.~Unmuth-Yockey, L.-P. Yang, A.~Bazavov, Z.~Y. Xie,
  T.~Xiang, S.~Chandrasekharan, S.-W. Tsai, and Y.~Meurice, {\it Progress
  towards quantum simulating the classical $\text{O}(2)$ model},  {\em Phys.
  Rev. A} {\bf 90} (Dec, 2014) 063603.

\end{thebibliography}
\end{document}